\pdfoutput=1
\documentclass[prl,letterpaper,twocolumn,preprintnumbers,aps,superscriptaddress,tightenlines,10pt]{revtex4-1}

\usepackage{amsmath,amssymb}
\usepackage{graphicx}

\newcommand{\nb}{\phantom{0}}

\begin{document}

\preprint{JLAB-THY-11-1428}

\title{Axial couplings and strong decay widths of heavy hadrons}

\author{William Detmold}
\affiliation{Department of Physics, College of William and Mary, Williamsburg, VA 23187, USA}
\affiliation{Jefferson Laboratory, 12000 Jefferson Avenue, Newport News, VA 23606, USA}
\author{C.-J.~David Lin}
\affiliation{Institute of Physics, National Chiao-Tung University, Hsinchu 300, Taiwan}
\affiliation{Physics Division, National Centre for Theoretical Sciences, Hsinchu 300, Taiwan}
\author{Stefan Meinel}
\affiliation{Department of Physics, College of William and Mary, Williamsburg, VA 23187, USA}

\begin{abstract}
We calculate the axial couplings of mesons and baryons containing a heavy quark in the static
limit using lattice QCD. These couplings determine the leading interactions in heavy hadron
chiral perturbation theory and are central quantities in heavy quark physics, as they control
strong decay widths and the light-quark mass dependence of heavy hadron observables. Our analysis
makes use of lattice data at six different pion masses, $227 {\:\:\rm MeV} < m_\pi < 352$ MeV,
two lattice spacings, $a=0.085$, 0.112 fm, and a volume of (2.7 fm)$^3$. Our results for the axial
couplings are $g_1=0.449(51)$, $g_2=0.84(20)$, and $g_3=0.71(13)$, where $g_1$ governs the
interaction between heavy-light mesons and pions and $g_{2,\,3}$ are similar couplings between
heavy-light baryons and pions. Using our lattice result for $g_3$, and constraining $1/m_Q$
corrections in the strong decay widths with experimental data for $\Sigma_c^{(*)}$ decays, we obtain
$\Gamma[\Sigma_b^{(*)} \!\!\to \!\Lambda_b\:\pi^\pm]=4.2(1.0),\, 4.8(1.1),\, 7.3(1.6),\, 7.8(1.8)$ MeV
for the $\Sigma_b^+$, $\Sigma_b^-$, $\Sigma_b^{*+}$, $\Sigma_b^{*-}$ initial states, respectively.
We also derive upper bounds on the widths of the $\Xi_b^{\prime(*)}$ baryons.
\end{abstract}

\maketitle

{\it Introduction.}---Significant progress has been made in the last few years in uncovering
the spectrum and decays of hadrons containing heavy quarks at the dedicated $B$ factories, the Tevatron,
and the LHC. Accurate lattice QCD calculations are required to confront data from these experiments with
the Standard Model. These lattice calculations involve extrapolations in the masses of the light quarks,
which require theoretical guidance. For hadrons containing a single heavy quark, the relevant effective theory
is known as heavy-hadron chiral perturbation theory (HH$\chi$PT)
\cite{Wise:1992hn, Burdman:1992gh, Yan:1992gz, Cho:1992cf}, which is built upon two of the most
important symmetries of QCD: chiral symmetry and heavy-quark symmetry. At leading order,
the HH$\chi$PT Lagrangian contains three \emph{axial couplings} $g_1$, $g_2$, and $g_3$.
The coupling $g_1$ determines the strength of the interaction between heavy-light mesons and pions, while
$g_2$ and $g_3$ similarly determine the interaction of heavy-light baryons with pions. Consequently,
these couplings are central to the low-energy dynamics of heavy-light hadrons, and can be used to calculate
the widths of strong decays such as $\Sigma_b^{(*)} \to \Lambda_b\:\pi$. The axial couplings are
calculable from the underlying theory of QCD, using a lattice regularization. The mesonic coupling $g_1$ has
been previously studied in lattice QCD with $N_f=0$ or $N_f=2$ dynamical quark flavors
\cite{deDivitiis:1998kj, Abada:2003un, Ohki:2008py, Becirevic:2009yb, Bulava:2010ej}.
In the following, we present the first complete calculation of $g_1$, $g_2$, and $g_3$ in $N_f=2+1$ lattice QCD,
controlling all systematic uncertainties. We use our results to calculate
$\Gamma[\Sigma_b^{(*)} \to \Lambda_b\:\pi^\pm]$
and give bounds on $\Gamma[\Xi_b^{\prime(*)} \to \Xi_b \:\pi]$. Technical details of the analysis
that are omitted here for brevity will be presented in a forthcoming paper.

{\it Lattice QCD calculation.}---The heavy hadrons considered in the lattice calculation are the
lowest-lying heavy-light mesons and baryons containing light valence quarks of the flavors $u$ or $d$.
We work in the heavy-quark limit $m_Q=\infty$ where the axial couplings are defined, and assume isospin
symmetry. The heavy-light mesons occur in degenerate pseudoscalar and vector multiplets, described by
interpolating fields $P^i\sim \bar{Q}\gamma_5 q^i$ and $P^{*i}_\mu \sim \bar{Q}\gamma_\mu q^i$,
where $q^i$ is a light quark of flavor $i$ and $\bar Q$ is a static heavy antiquark. In the heavy-light
baryon sector, we include both the states with $s_l=0$ and $s_l=1$, where $s_l$ is the (conserved)
spin of the light degrees of freedom. The states with $s_l=1$ are described by an interpolating field
$S^{ij}_{\mu\:\alpha} \sim \epsilon_{abc}\:(C\gamma_\mu)_{\beta\gamma}\:q^i_{a\beta}\:q^j_{b\gamma}\: Q_{c\alpha}$
that couples to the isotriplet states with both $J=1/2$ ($\Sigma_Q$) and $J=3/2$ ($\Sigma_Q^*$), which are
degenerate in the heavy-quark limit. The isosinglet $s_l=0$ baryon $\Lambda_Q$ has $J=1/2$ and is described
by an interpolating field
$T^{ij}_\alpha\sim \epsilon_{abc}\:(C\gamma_5)_{\beta\gamma}\:q^i_{a\beta}\:q^j_{b\gamma}\: Q_{c\alpha}$.
The axial couplings can be extracted by calculating matrix elements of the axial current
$A_\mu\sim\bar{d}\gamma_\mu\gamma_5 u$:
\begin{eqnarray}
  \nonumber  \langle P^*_d  | A_\mu | P_u  \rangle &=& -2\:(g_1)_{\rm eff} \: \varepsilon^*_\mu, \\
  \nonumber  \langle S_{dd} | A_\mu | S_{du} \rangle &=& -(i/\sqrt{2})\:(g_2)_{\rm eff} \:v^\sigma
              \: \epsilon_{\sigma \mu \nu\rho}\: \overline{U}^\nu U^\rho, \\
  \langle S_{dd} | A_\mu | T_{du} \rangle &=& -(g_3)_{\rm eff} \:  \overline{U}_\mu \: \mathcal{U}.
  \label{eq:LOmatrixelts}
\end{eqnarray}
Here, $v$ is the four-velocity, $\varepsilon^\mu$ is the polarization vector of the $P^*$ meson,
$\mathcal{U}$ is the Dirac spinor of the $T$ baryon, and the $U^\mu$'s are the ``superfield
spinors'' of the $S$ baryons \cite{Detmold:2011rb}. At leading order in the chiral expansion,
the ``effective axial couplings'' $(g_i)_{\rm eff}$ defined via (\ref{eq:LOmatrixelts}) are equal
to the axial couplings $g_i$ that appear in the HH$\chi$PT Lagrangian. The next-to-leading-order expressions for
$(g_i)_{\rm eff}$ are given in Ref.~\cite{Detmold:2011rb}. To calculate the matrix elements
(\ref{eq:LOmatrixelts}) in lattice QCD, we set $\mathbf{v}=0$ and construct Euclidean two- and
three-point correlators $C_H(t)=\langle \chi_H({\bf x},t)\:\chi^\dag_H({\bf x},0) \rangle$ and
$C_{H \to H^\prime}(t,t')={\sum_{\bf x'}}\langle \chi_{H^\prime} ({\bf x},t)
\: A_\mu({\bf x'},t') \: \chi_H^\dag({\bf x},0) \rangle$, where $t>t'>0$ and $\chi_H$ are the
interpolating fields of the heavy hadrons as defined above. We form the ratios
\begin{eqnarray}
  R_1(t, t') &=&  -\frac{\frac13 C_{P_u\to P_d^{*}}^{\mu\mu}(t, t')}{C_{P_u}(t)}, \label{eq:R1}
\end{eqnarray}
\begin{eqnarray}
  R_2(t, t') &=& 2\frac{\frac i6 \epsilon_{0\mu\nu\rho}\: C_{S_{du}\to S_{dd}}^{\mu\nu\rho}(t, t') }
  { \frac13 C_{S_{dd}}^{\mu\mu}(t) }, \label{eq:R2}
\end{eqnarray}
and the double ratio (needed because of the nonzero $S-T$ mass splitting)
\begin{equation}
  R_3(t, t') = \sqrt{\frac{ \frac13 C_{T_{du}\to S_{dd}}^{\mu\mu}(t, t')
  \:\: \frac13 C_{S_{dd}\to T_{du}}^{\nu\nu}(t, t')  }
  { \frac13 C_{S_{dd}}^{\mu\mu}(t) \: C_{T_{du}}(t) }}. \label{eq:R3}
\end{equation}
Here, $\mu,\nu,\rho$ are the Lorentz indices from the axial current or the interpolating fields
for $P^*$ and $S$ and are summed over when repeated. Using (\ref{eq:LOmatrixelts}) and the
spectral decomposition of the correlators, one finds that
\begin{equation}
  R_i(t, t/2) = (g_i)_{\rm eff} + O(e^{-\delta_i t}), \label{eq:geffR}
\end{equation}
where the $\delta_i$ are related to the energy gaps of the lowest contributing excited states.

The calculations presented in this work make use of lattice gauge field configurations
generated by the RBC/UKQCD collaboration \cite{Aoki:2010dy} with $2+1$ flavors of light quarks,
implemented with a domain-wall action that realizes lattice chiral symmetry. The details of the
ensembles included in our analysis can be found in Table \ref{tab:params}. We computed
domain-wall light-quark propagators for a range of unitary ($am_{u,d}^{(\mathrm{val})}=am_{u,d}^{(\mathrm{sea})}$)
and partially quenched ($am_{u,d}^{(\mathrm{val})}<am_{u,d}^{(\mathrm{sea})}$) quark masses.
As shown in the lower part of the table, we have data with (valence) pion masses ranging from
227 to 352 MeV, two lattice spacings, $a=0.085$, 0.112 fm, and a large lattice volume
of (2.7 fm)$^3$. The sea-strange-quark masses are about 10\% above the physical value,
and we assign a 1.5\% systematic uncertainty to our final results to account for this,
based on the size of the effect on similar observables as studied in Ref.~\cite{Aoki:2010dy}.
For the light-quark propagators, we used gauge-invariant Gaussian smeared sources to
improve the overlap of the hadron interpolating fields with the ground states.
We constructed the three-point functions $C_{H \to H^\prime}(t,t')$ using light-quark
propagators with smeared sources at $({\bf x},0)$ and $({\bf x},t)$ and a local sink at the current
insertion point $({\bf x'},t')$, for various separations $t$ as shown in Table \ref{tab:params}.
The bare lattice axial current requires a finite renormalization $Z_A$ to match the continuum current,
$A_\mu = Z_A \:\overline{u}\gamma_\mu\gamma_5 d$. We used nonperturbative results for $Z_A$ obtained
by the RBC/UKQCD collaboration \cite{Aoki:2010dy}.

The action for the static heavy quark is a modified form of the Eichten-Hill action \cite{Eichten:1989kb}
in which the standard gauge links are replaced by HYP (hypercubic) smeared \cite{Hasenfratz:2001hp} gauge links,
resulting in improved statistical signals for the correlators \cite{DellaMorte:2003mn}.
To study heavy-quark discretization effects and optimize the signals, we generated data for
$n_{\rm HYP}=1,2,3,5,10$ levels of HYP smearing, corresponding to different lattice actions
for the heavy quarks. These actions have the same continuum limit, but may scale differently.
Our final analysis focuses on $n_{\rm HYP}=1,2,3$.

\begin{table}[t]
  \centering
  \begin{ruledtabular}
  \begin{tabular}{cccccccccc}
  Ensemble &  $a$ (fm) 	& & $L^3\times T$  & & $am_{u,d}^{(\mathrm{sea})}$ &  & $m_\pi^{(\mathrm{ss})}$ (MeV)   \\
  \hline
  A & 0.1119(17) && $24^3\times64$  && 0.005 && 336(5) \\ 
  B & 0.0849(12) && $32^3\times64$  && 0.004 && 295(4) \\
  C & 0.0848(17) && $32^3\times64$  && 0.006 && 352(7) \\
  \end{tabular}
  \vspace*{-1.2mm}
  \begin{tabular}{ccccccc}
  Ensemble &  $am_{u,d}^{(\mathrm{val})}$  & $m^{(\rm vs)}_\pi$ (MeV) & $m^{(\rm vv)}_\pi$ (MeV) & $t/a$ \\
  \hline
  A &  0.001 & 294(5) & 245(4) & 4, 5, ..., 10 \\
  A &  0.002 & 304(5) & 270(4) & 4, 5, ..., 10 \\
  A &  0.005 & 336(5) & 336(5) & 4, 5, ..., 10 \\
  B &  0.002 & 263(4) & 227(3) & 6, 9, 12 \\
  B &  0.004 & 295(4) & 295(4) & 6, 9, 12 \\
  C &  0.006 & 352(7) & 352(7) & 13 \\
  \end{tabular}
  \end{ruledtabular}
  \caption{\label{tab:params}Details of gauge field ensembles (upper section, see also Ref.~\cite{Aoki:2010dy})
  and ``measurements'' (lower section). The superscripts $v$, $s$ on $m_{\pi}$ indicate the masses of the
  quarks in the pions, equal to $am_{u,d}^{(\mathrm{val})}$ or $am_{u,d}^{(\mathrm{sea})}$.}
\end{table}

\begin{figure}[t]
  \includegraphics[width=0.83\linewidth]{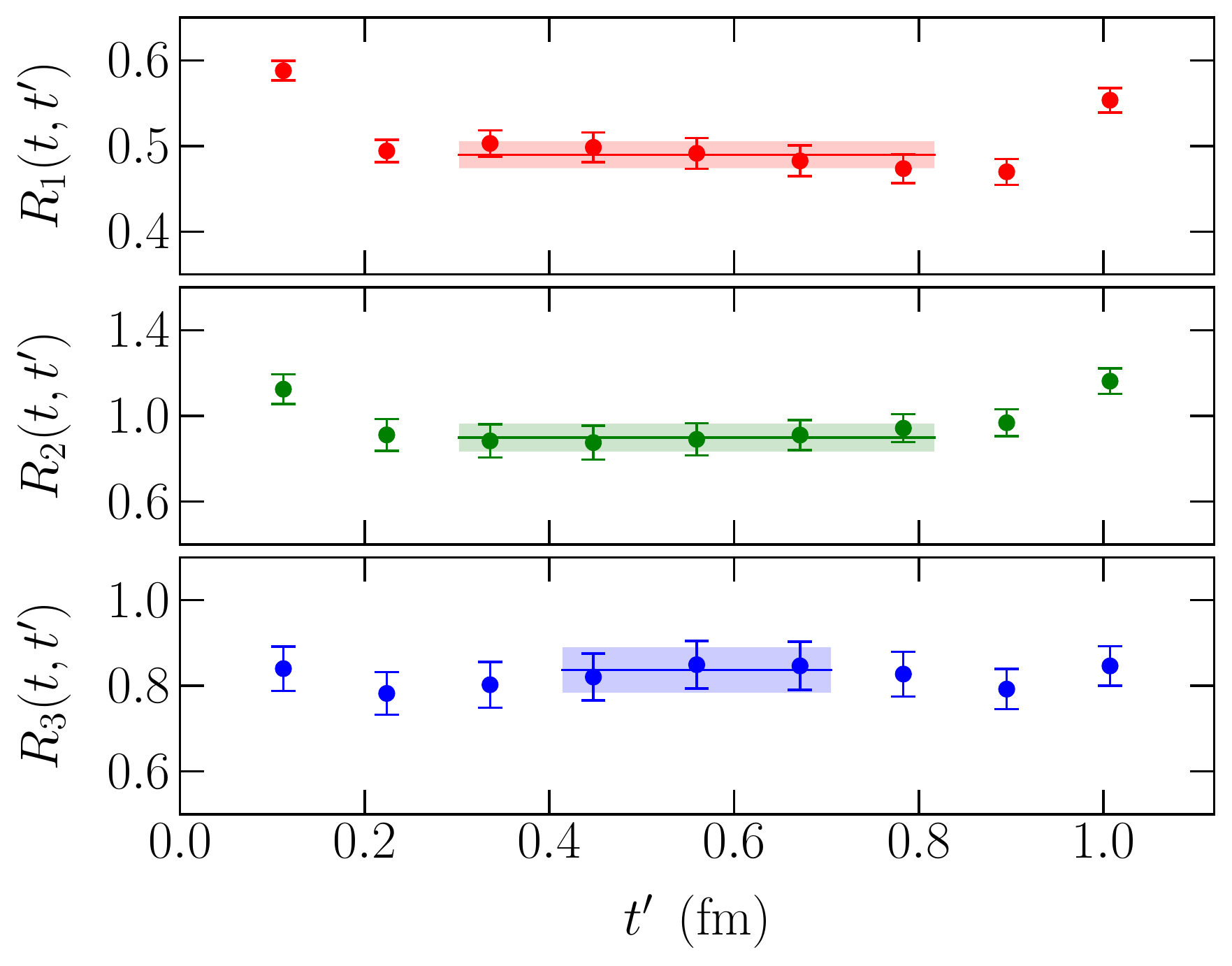}
  \caption{\label{fig:ratios}Ratios $R_i(t,t')$ as a function of the current insertion time slice $t'$, for $t/a=10$,
  at $a=0.112$ fm, $am_{u,d}^{(\mathrm{val})}=0.002$, $n_{\rm HYP}=3$.}
\end{figure}

\begin{figure}[t]
  \includegraphics[width=0.9\linewidth]{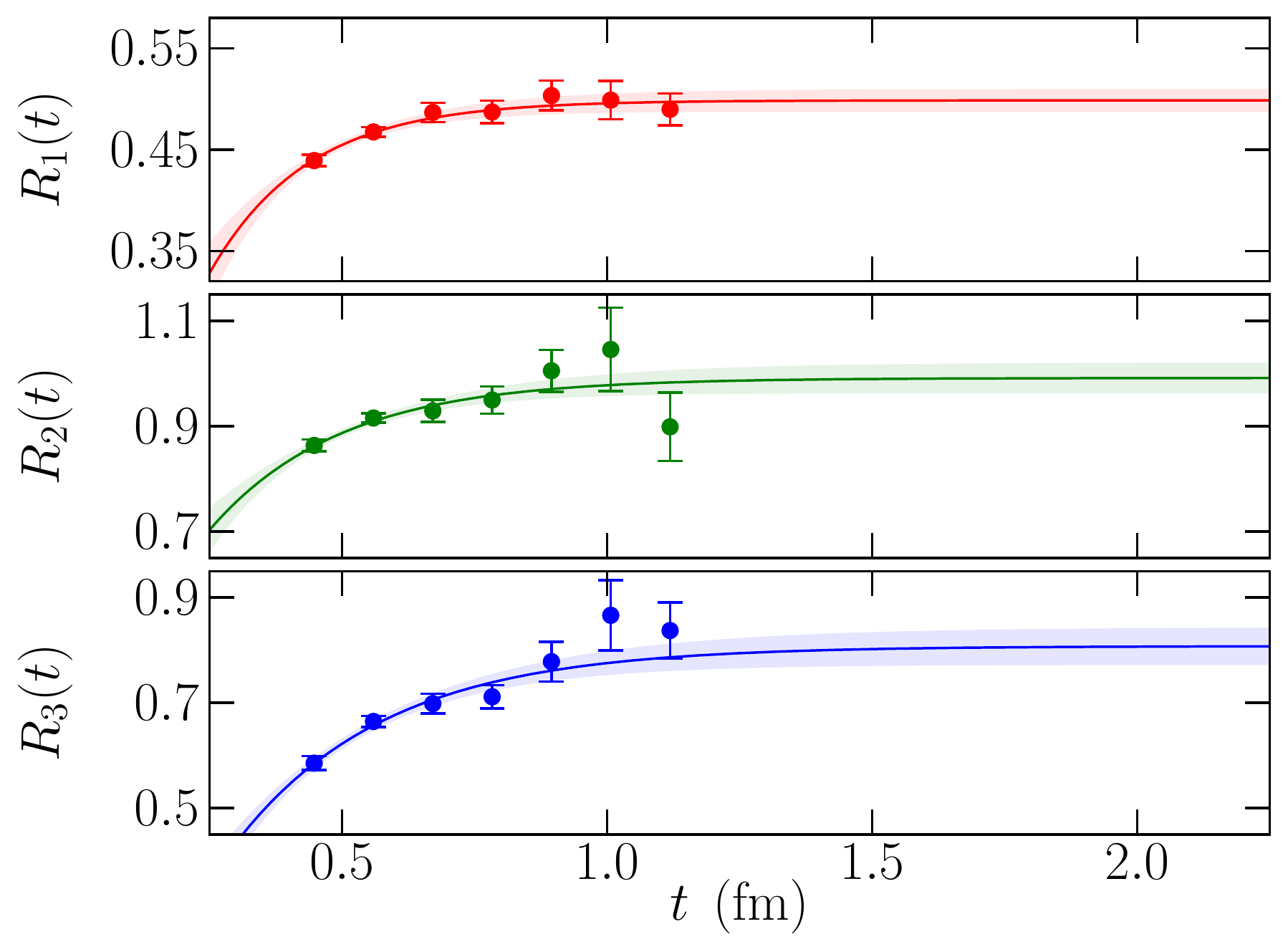}
  \caption{\label{fig:tsepextrapolation}Fits of the $t$-dependence of $R_i(t)$, for $a=0.112$ fm,
  $am_{u,d}^{(\mathrm{val})}=0.002$, $n_{\rm HYP}=3$.}
\end{figure}

In Fig.~\ref{fig:ratios}, we show examples of numerical results for the ratios (\ref{eq:R1}),
(\ref{eq:R2}), and (\ref{eq:R3}). We observed plateaus in $R_i(t,\: t')$ as a function of $t'$,
and we averaged the ratios in this region, which is essentially equivalent to taking $R_i(t, t/2)$.
We denote these averages as $R_i(t)$. To obtain the ground-state contributions according to (\ref{eq:geffR}),
one needs to calculate $\lim_{t\rightarrow\infty} R_i(t)$. To this end, we performed fits of the data using
the functional form $R_i(t)=(g_i)_{\rm eff}-A_i \:e^{-\delta_i\:t}$ with parameters $(g_i)_{\rm eff}$,
$A_i$ and $\delta_i$, depending on the lattice spacing $a$, the quark masses $am_{u,d}^{(\mathrm{val})}$,
$am_{u,d}^{(\mathrm{sea})}$, and the smearing parameter $n_{\rm HYP}$. This functional form only
includes the leading contributions from excited states, but was able to fit the data well, as shown in
Fig.~\ref{fig:tsepextrapolation}.
We used the results and uncertainties for the gap parameters $\delta_i$
from the fits at the coarse lattice spacing to constrain the fits at the fine lattice spacing, where we have
fewer values of $t/a$. As explained in Ref.~\cite{Detmold:2012ge}, we then additionally constrained the parameters $A_i$
(independently for the two different lattice spacings),
using information from initial fits of data from ensembles A and B. This allowed us to perform fits using the same form
of the function $R_i(t)$ even for the data from ensemble C, where we have only one value of $t/a$.
To estimate the systematic uncertainties caused by higher excited states, we calculated
the shifts in $(g_i)_{\rm eff}$ at the coarse lattice spacing when removing one or two data points with the smallest
$t/a$ ($=4,5$) or adding a second exponential to the fits \cite{Detmold:2012ge}. Repeated fits
of $R_i(t)$ for a bootstrap ensemble allowed the calculation of the covariance matrices describing the
correlations of the results for $(g_i)_{\rm eff}$ from common ensembles of gauge field configurations.

Having obtained the results for $(g_i)_{\rm eff}$, we then performed fully correlated fits of the 
$a$-, $m_\pi^{(\rm vv)}$-, and $m_\pi^{(\rm vs)}$-dependence. For $(g_1)_{\rm eff}$, we used the function
\begin{eqnarray}
  \nonumber (g_1)_{\mathrm{eff}} &=& g_1\Big[1 + f_1(g_1,m_\pi^{(\mathrm{vv})}, m_\pi^{(\mathrm{vs})}, L)
  + d_{1,n_\mathrm{HYP}}\:a^2  \\
  && \phantom{g_1\Big[1} + c_1^{(\mathrm{vv})} \:[m_\pi^{(\mathrm{vv})}]^2 + c_1^{(\mathrm{vs})}\:
  [m_\pi^{(\mathrm{vs})}]^2 \:\Big], \label{eq:g1fitfunc}
\end{eqnarray}
where $g_1$, $c_1^{(\mathrm{vv})}$, $c_1^{(\mathrm{vs})}$, $\{ d_{1,n_\mathrm{HYP}} \}$ are the free parameters.
For $(g_2)_{\rm eff}$ and $(g_3)_{\rm eff}$, we performed coupled fits using
\begin{eqnarray}
  \nonumber (g_i)_{\mathrm{eff}} &=& g_i\Big[1 + f_i(g_2, g_3, m_\pi^{(\mathrm{vv})},
  m_\pi^{(\mathrm{vs})}, \Delta, L) + d_{i,n_\mathrm{HYP}}\:a^2 \\
  && \phantom{g_i\Big[1}+ c_i^{(\mathrm{vv})} \: [m_\pi^{(\mathrm{vv})}]^2
  + c_i^{(\mathrm{vs})}\: [m_\pi^{(\mathrm{vs})}]^2 \: \Big]\phantom{X} \label{eq:g2g3fitfunc}
\end{eqnarray}
(for $i=2,3$), where the free fit parameters are $g_2$, $g_3$, $c_2^{(\mathrm{vv})}$, $c_3^{(\mathrm{vv})}$,
$c_2^{(\mathrm{vs})}$, $c_3^{(\mathrm{vs})}$, $\{ d_{2,n_\mathrm{HYP}}, d_{3,n_\mathrm{HYP}} \}$.
The functions $f_i$ in (\ref{eq:g1fitfunc}) and (\ref{eq:g2g3fitfunc}) are the nonanalytic loop contributions
in partially quenched $SU(4|2)$ HH$\chi$PT and can be found in Ref.~\cite{Detmold:2011rb}. They also include
the leading effects of the finite lattice size $L$ (because of our large volume, the finite-volume corrections
were smaller than 3\% for all data points). The functions $f_i$ depend on the renormalization scale $\mu$,
but this dependence is canceled exactly by the $\mu$-dependence of the counterterms $c_i^{(\mathrm{vv})}$
and $c_i^{(\mathrm{vs})}$. The parameters $d_{i,n_\mathrm{HYP}}$ for each $n_{\rm HYP}$ describe the leading
effects of the nonzero lattice spacing, which are multiplicative corrections proportional to $a^2$ as a
consequence of the lattice chiral symmetry of the domain-wall action. In (\ref{eq:g2g3fitfunc}),
the quantity $\Delta$ is the $S-T$ mass splitting, which we set to $\Delta=200$ MeV in our fits, consistent
with experiments \cite{Aaltonen:2007rw, Tonelli:2010tz} and our lattice data (note that $\Delta$
does not vanish in the chiral or heavy-quark limits).

\begin{figure}[t]
  \includegraphics[width=0.9\linewidth]{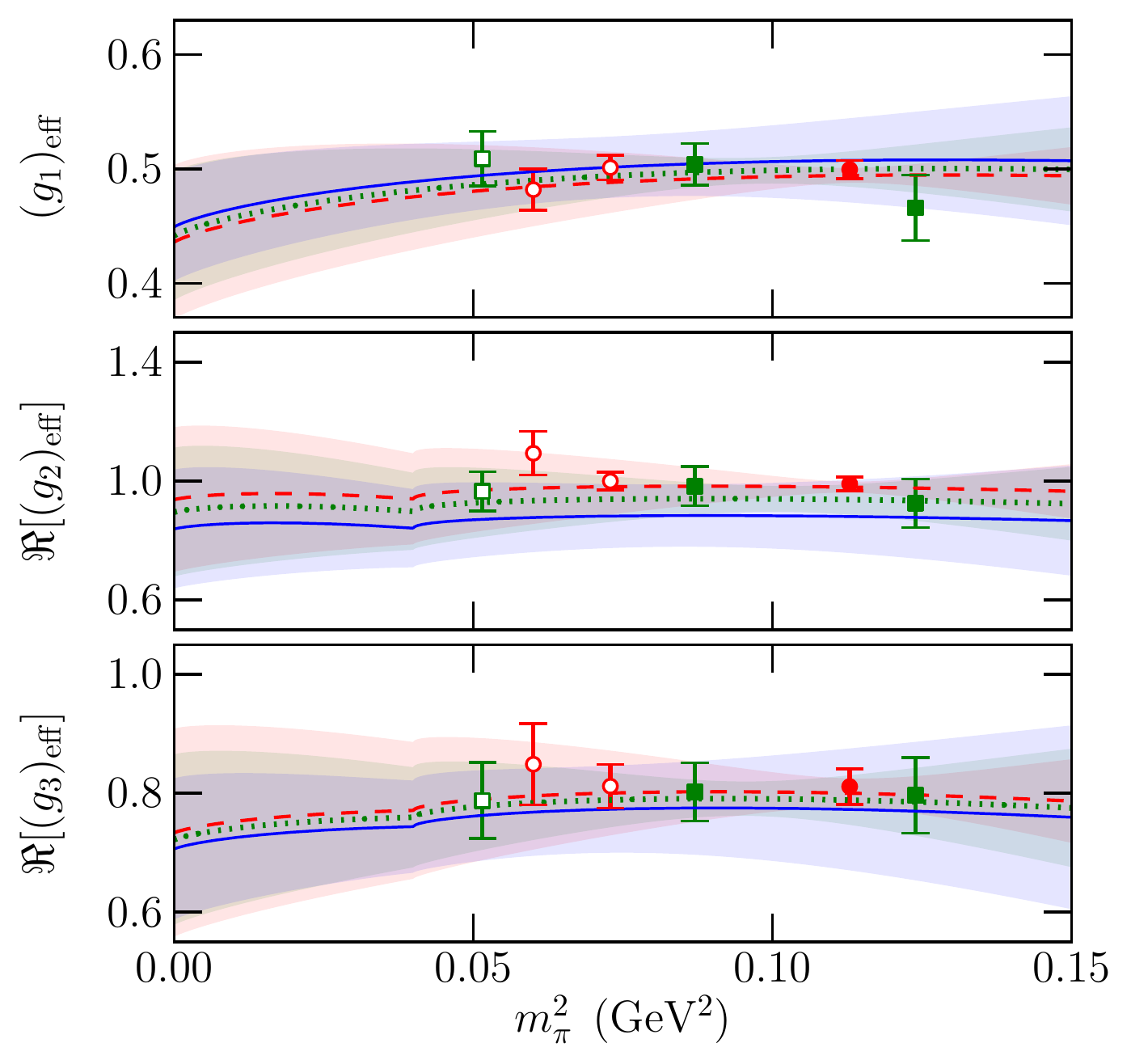}
  \caption{\label{fig:madep}The (real parts of the) fitted functions
  $(g_1)_{\rm eff}$, $(g_2)_{\rm eff}$, $(g_3)_{\rm eff}$, evaluated in infinite volume
  and $n_{\rm HYP}=3$, for the unitary case $m_\pi^{(\mathrm{vv})}=m_\pi^{(\mathrm{vs})}=m_\pi$.
  The dashed line corresponds to $a=0.112$ fm, the dotted line to $a=0.085$ fm, and the solid line
  to the continuum limit. The shaded regions indicate the $1\sigma$ statistical uncertainty.
  Also shown are the data points, shifted to infinite volume (circles: $a=0.112$ fm, squares:
  $a=0.085$ fm). The partially quenched data points (open symbols), which have
  $m_\pi^{(\mathrm{vv})}<m_\pi^{(\mathrm{vs})}$, are included in the plot at $m_\pi=m_\pi^{(\mathrm{vv})}$,
  even though the fit functions actually have slightly different values for these points.}
\end{figure}

To determine for which values of $n_{\rm HYP}$ the order-$a^2$ corrections in (\ref{eq:g1fitfunc})
and (\ref{eq:g2g3fitfunc}) adequately describe the lattice artefacts in the data, we started from
fits that included all values of $n_{\rm HYP}$, and then successively removed the data with the largest
values of $n_{\rm HYP}$. After excluding $n_{\rm HYP}=10$ and $n_{\rm HYP}=5$, we obtained good
quality-of-fit values [$Q=0.70$ for $(g_1)_{\rm eff}$ and $Q=0.92$ for $(g_{2,3})_{\rm eff}$],
and the results were stable under further exclusions. Our final results for the axial couplings,
taken from the fits with $n_{\rm HYP}=1,2,3$, are
\begin{eqnarray}
  g_1&=&0.449   \pm 0.047_{\:\rm stat}   \pm 0.019_{\:\rm syst}, \nonumber \\
  g_2&=&0.84\nb \pm 0.20_{\:\rm stat}\nb \pm 0.04_{\:\rm syst},  \nonumber \\
  g_3&=&0.71\nb \pm 0.12_{\:\rm stat}\nb \pm 0.04_{\:\rm syst}. \label{eq:finalresults}
\end{eqnarray}
Independent fits for each $n_{\rm HYP}$ (1, 2, 3, 5, 10) gave results consistent
with (\ref{eq:finalresults}). The estimates of the systematic uncertainties
in (\ref{eq:finalresults}) include the following \cite{Detmold:2012ge}: effects of next-to-next-to-leading-order terms in the fits
to the $a$- and $m_\pi$-dependence (3.6\%, 2.8\%, 3.7\% for $g_1$, $g_2$, $g_3$, respectively),
effects from the unphysically large sea-strange-quark mass (1.5\%), and effects from higher excited states
in the $t\to\infty$ extrapolations of $R_i(t)$ (1.7\%, 2.8\%, 4.9\%). The resulting mass- and
lattice-spacing dependence of the effective couplings from the fits with (\ref{eq:g1fitfunc}) and
(\ref{eq:g2g3fitfunc}) is shown in Fig.~\ref{fig:madep}. Note that the functions $(g_2)_{\rm eff}$
and $(g_3)_{\rm eff}$ develop small imaginary parts for pion masses below the $S\to T \pi$
\mbox{threshold} at $m_\pi = \Delta$ \cite{Detmold:2011rb} (the lattice data are all above this \mbox{threshold}),
and the real parts are shown in the figure. The fitted coefficients $d_{i,\:n_{\rm HYP}}$ are consistent
with zero within statistical uncertainties, and the analytic counterterms $c_i^{(\mathrm{vv})}$ and
$c_i^{(\mathrm{vs})}$ are natural-sized (when evaluated at $\mu=4\pi f_\pi$ with $f_\pi = 132$ MeV),
indicating that the chiral expansions of $(g_i)_{\mathrm{eff}}$ are under control for the range of
masses used here.

\begin{figure}[t!]
  \includegraphics[width=0.86\linewidth]{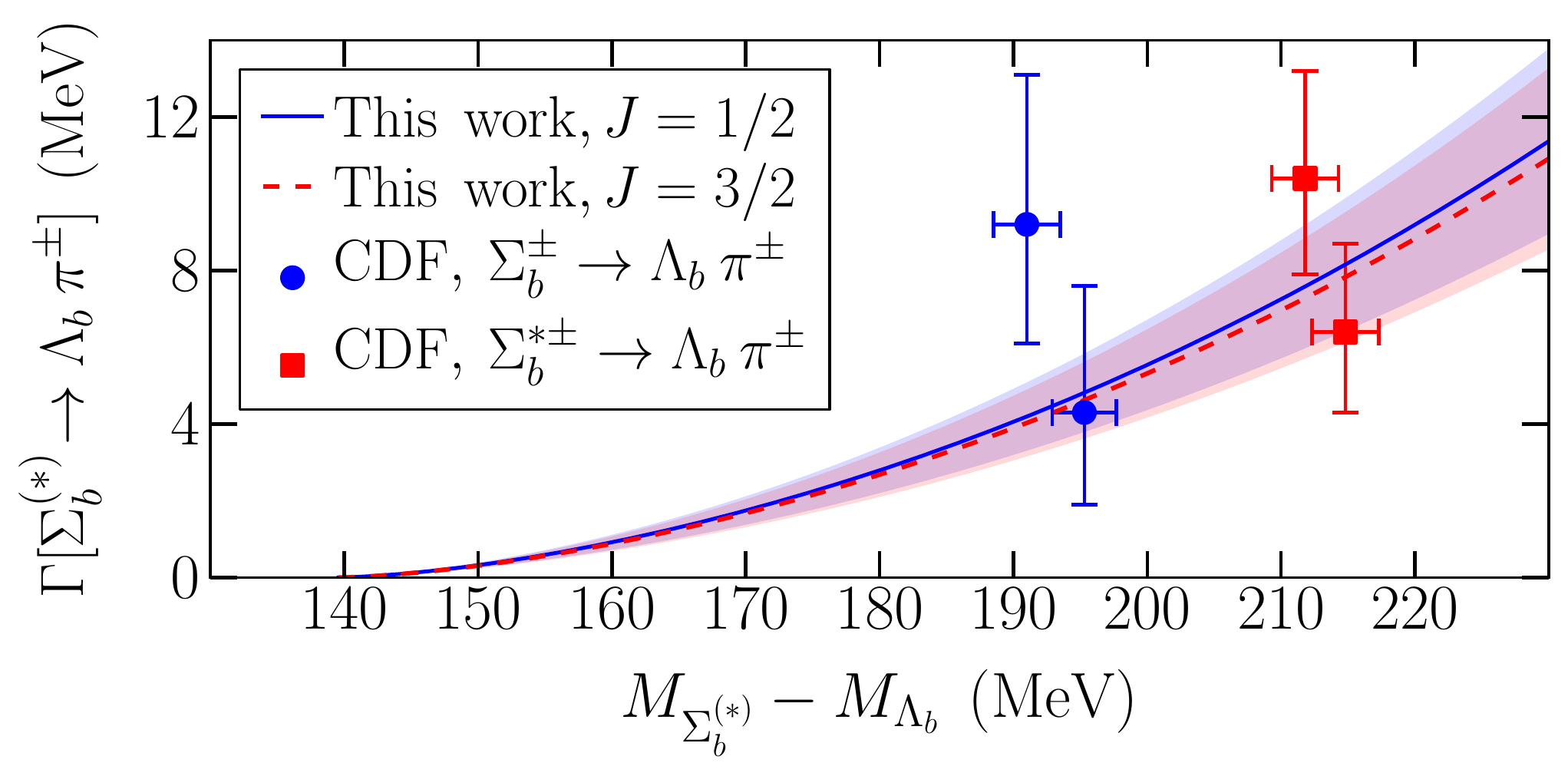}
  \caption{\label{fig:strongdecay}Widths of the decays
  $\Sigma_b^{(*)\pm}\!\to \Lambda_b\:\pi^\pm$ as functions
  of the $\Sigma_b^{(*)}-\Lambda_b$ mass difference. The curves (solid: $\Sigma_b$, dashed: $\Sigma_b^*$)
  and shaded regions show our predictions and their uncertainties.
  The data points are from CDF \cite{Tonelli:2010tz}.}
\end{figure}

{\it Calculation of strong decay widths.}---At leading order in the chiral expansion,
the widths for the strong decays $S\to T\:\pi$ are
\begin{equation}
  \Gamma[S \to T\: \pi]=c_{\rm f}^2\:\frac{1}{6\pi f_\pi^2}
  \left(g_3+\frac{\kappa_J}{m_Q}\right)^2\frac{M_T}{M_S} \: |\mathbf{p}_\pi|^3,
  \label{eq:decaywidth}
\end{equation}
where $S$ and $T$ now denote physical $s_l=1$ and $s_l=0$ heavy baryon states such as $\Sigma_b$
and $\Lambda_b$, $|\mathbf{p}_\pi|$ is the magnitude of the pion momentum in the $S$ rest frame,
and $c_{\rm f}$ is a flavor factor, equal to 1 for $\Sigma_Q^{(*)} \!\to \Lambda_Q \:\pi^\pm$,
$1/\sqrt{2}$ for $\Xi_Q^{\prime(*)} \!\to \Xi_Q \: \pi^\pm$, and $1/2$ for $\Xi_Q^{\prime(*)}\! \to \Xi_Q \: \pi^0$.
Here we modified the $m_Q=\infty$
expression for $\Gamma$ \cite{Pirjol:1997nh} by including the term $\kappa_J/m_Q$. Terms suppressed
by $(m_\pi/\Lambda_\chi)^2$ and $(\Lambda_{\rm QCD}/m_Q)^2$, which are omitted from
(\ref{eq:decaywidth}), lead to small systematic uncertainties in $\Gamma$. To determine $\kappa_{1/2}$
and $\kappa_{3/2}$, we performed fits of experimental data \cite{Nakamura:2010zzi}
for the widths of the $\Sigma_c^{++}$, $\Sigma_c^0$ ($J=1/2$)
and the $\Sigma_c^{*++}$, $\Sigma_c^{*0}$ ($J=3/2$) using (\ref{eq:decaywidth}),
where we constrained $g_3$ to our lattice QCD result (\ref{eq:finalresults}) and set $m_Q=\frac12 M_{J/\psi}$. These fits
gave $\kappa_{1/2}=0.55(21)$ GeV and $\kappa_{3/2}=0.47(21)$ GeV.
We then evaluated (\ref{eq:decaywidth}) for $m_Q=\frac12 M_\Upsilon$ to obtain
predictions for the decays of bottom baryons. Our calculated widths
$\Gamma[\Sigma_b^{(*)}\! \to \Lambda_b\:\pi^\pm]$ as functions of the
$\Sigma_b^{(*)}-\Lambda_b$ mass difference are shown as the curves in Fig.~\ref{fig:strongdecay}.
Using the experimental values of the baryon masses \cite{Tonelli:2010tz, Nakamura:2010zzi},
our results for $\Gamma[\Sigma_b^{(*)}\! \to \Lambda_b\:\pi^\pm]$
in MeV are 4.2(1.0), 4.8(1.1), 7.3(1.6), 7.8(1.8) for the $\Sigma_b^+$, $\Sigma_b^-$, $\Sigma_b^{*+}$,
$\Sigma_b^{*-}$ initial states, respectively, in agreement with the widths measured by the CDF
collaboration \cite{Tonelli:2010tz}. The decays $\Xi_b^{\prime(*)-}\!\! \to \Xi_b^-\pi^0,\: \Xi_b^0\,\pi^-$
and $\Xi_b^{\prime(*)0} \to \Xi_b^-\pi^+,\: \Xi_b^0\:\pi^0$ may
also be allowed, depending on the mass differences. With a spin-averaged
$\Xi_b^{\prime(*)}-\Xi_b$ splitting of $153(21)$ MeV (based on lattice data from
Ref.~\cite{Lewis:2008fu}), and assuming
$M(\Xi_b^{*})-M(\Xi_b^{\prime})\approx M(\Sigma_b^*)-M(\Sigma_b)$ = 21(2) MeV
\cite{Aaltonen:2007rw}, we obtain upper bounds of 1.1 and 2.8 MeV (CL=90\%)
for the total widths of the $\Xi_b^{\prime}$ and $\Xi_b^{*}$, respectively.

{\it Conclusions.}---We have presented a lattice QCD calculation of the axial couplings
of hadrons containing a heavy quark in the static limit, including for the first time
the baryonic couplings. We have used these results to predict the strong decay widths
of bottom baryons. Our calculation of the axial couplings controls all systematic
uncertainties by using two different lattice spacings, low pion masses, a large volume,
and the correct next-to-leading-order expressions from HH$\chi$PT. Since the axial couplings
are essential for chiral extrapolations of lattice data, their accurate determination is of
broad significance in flavor physics phenomenology.

{\it Acknowledgments.}---We thank H.-Y.~Cheng, K. Orginos, B.~Tiburzi, A.~Walker-Loud, and M.~Wingate
for discussions, R.~Edwards and B.~Jo\'o for the development of the {\tt chroma} library, and the
RBC/UKQCD collaboration for providing the gauge field configurations. The work of WD is supported
in part by JSA, LLC under DOE Contract No.~DE-AC05-06OR-23177 and by the Jeffress Memorial Trust,
J-968. WD and SM were supported by DOE OJI Award D{E}-S{C}000-1784 and DOE Grant No.~DE-FG02-04ER41302.
CJDL is supported by NSC Grant No.~99-2112-M-009-004-MY3.
We acknowledge the hospitality of Academia Sinica Taipei and NCTS Taiwan. This research
made use of computational resources provided by NERSC and the NSF Teragrid.

\end{document}